\documentclass[sigconf]{acmart}

\usepackage{enumitem}
\usepackage{graphicx}
\usepackage{soul}
\AtBeginDocument{%
  \providecommand\BibTeX{{%
    \normalfont B\kern-0.5em{\scshape i\kern-0.25em b}\kern-0.8em\TeX}}}

\copyrightyear{2023}
\acmYear{2023}
\setcopyright{rightsretained}
\acmConference[IPSN '23]{The 22nd International Conference on Information Processing in Sensor Networks}{May 9--12, 2023}{San Antonio, TX, USA}
\acmBooktitle{The 22nd International Conference on Information Processing in Sensor Networks (IPSN '23), May 9--12, 2023, San Antonio, TX, USA}
\acmDOI{10.1145/3583120.3589833}
\acmISBN{979-8-4007-0118-4/23/05}

\begin{document}

\title{Poster Abstract: Activity-based Profiling for Energy Harvesting Estimation}





 \author{Jiajie Li, Amani Abusafia, Abdallah Lakhdari, and Athman Bouguettaya} 
    \affiliation{ 
    \institution{The University of Sydney, Australia}
  \country{Australia}
    }
    \email{{jiajie.li, amani.abusafia, abdallah.lakhdari, athman.bouguettaya}@sydney.edu.au }

\renewcommand{\shortauthors}{Jiajie and Amani,  et al.}

\begin{abstract}
We propose a novel activity-based profiling framework to estimate IoT users' harvested energy based on their daily activities. Energy is harvested from natural sources such as the kinetic movement of IoT users. The profiling framework captures the users' physical activity data to define activity-based profiles. These profiles are utilized to estimate the harvested energy by IoT users. We train and evaluate our framework based on a real Fitbit dataset.

\end{abstract}

\ccsdesc[500]{Human-centered computing~Ubiquitous computing}
\ccsdesc[500]{Human-centered computing~Ubiquitous and mobile computing systems and tools}

\keywords{Wireless energy, IoT,   Energy Harvesting, activity profiling, wireless power transfer, energy services}

\maketitle
\vspace{-5pt}
\section{Introduction}

\emph{ Energy services} refers to the \textit{wireless transfer of energy}  from an IoT device (e.g. \textit{energy provider}) to a nearby device (e.g., \textit{energy consumer}) \cite{lakhdari2020Vision}\cite{abusafia2022services}  Energy providers, such as smart shoes, may \textit{harvest} energy from natural resources (e.g., physical activity) \cite{sandhu2020phd}\cite{gorlatova2015movers} For example, PowerWalk kinetic energy harvester  produces 10-12 watts of on-the-move power\cite{lakhdari2020Vision}. The harvested energy may be used to charge nearby IoT devices wirelessly using new technologies known as \textit{``Over-the-Air wireless charging''} \cite{abusafia2022maximizing}\cite{Amani2022QoE}\cite{yang2023Monitoring}. For example, Energous may charge devices up to a distance of 4.5 meters \cite{yao2022wireless}\cite{yang2022towards}.\looseness=-1

Enabling a sustainable energy services IoT ecosystem will offer ubiquitous access to energy \cite{abusafia2022services}. In this environment, IoT devices can be charged automatically  without human intervention. To achieve that, the system must know \textit{when}, \textit{where}, and \textit{how much} energy an IoT device may request/offer as a service from/to nearby IoT devices \cite{lakhdari2020Vision}. This paper focuses on estimating the energy services that a provider may offer, specifically on \textit{estimating how much energy a provider may harvest to offer as a service}.\looseness=-1

\begin{figure}[!t]
    \centering
    \setlength{\abovecaptionskip}{5pt}
    \setlength{\belowcaptionskip}{-15pt}
    \includegraphics[width=\linewidth]{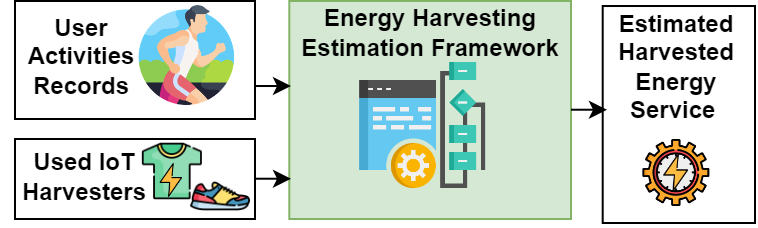}
    \caption{Energy Harvesting Estimation System}
    \label{general}
\end{figure}

Studies have proposed to estimate the energy from daily activities by measuring the output from kinetic energy harvesters \cite{sandhu2020phd}. Harvested energy may be estimated by capturing the kinetic movement of the device wearer and their body temperature. Moreover, other studies used kinetic energy harvesting data  to recognize daily activities \cite{khalifa2017harke}\cite{khalifa2015energy} In principle, it is possible to estimate the amount of harvested energy based on the activity and the worn harvester's type \cite{khalifa2017harke} Such activity may be collected by wearables such as smartwatches. In this paper, we envision  a novel framework to profile users based on their daily physical activities, consisting of an activity-based profiling component and an energy-harvesting estimation model (See Fig. \ref{general}) \cite{lakhdari2020Vision}.  The activity-based profiling component will aim to classify users based on their daily activities in order to associate them with the corresponding energy harvesting estimation model. The  model will use the activities profiles and the type of worn harvester to estimate the harvested energy.  As energy harvesting technologies are still developing,\textit{we focus on the activity-based profiling}.\looseness=-1

\begin{table*}[!t]
\setlength{\abovecaptionskip}{5pt}
\setlength{\belowcaptionskip}{-10pt}
\centering
\caption{Evaluation Metrics of clustering algorithms}
\label{resultstable}

\resizebox{12cm}{!} 
{ 
\begin{tabular}{l|ccc|ccc|}
\cline{2-7}
\multicolumn{1}{c|}{}                                 & \multicolumn{3}{c|}{\textbf{K-means Clustering}}                                                                            & \multicolumn{3}{c|}{\textbf{Robust Clustering}}                                                                             \\ \hline
\multicolumn{1}{|l|}{\textbf{ Metrics $\backslash$ Feature Selection}}     & \multicolumn{1}{c|}{\textbf{Original}} & \multicolumn{1}{c|}{\textbf{PCA}} & \textbf{Correlation} & \multicolumn{1}{c|}{\textbf{Original}} & \multicolumn{1}{c|}{\textbf{PCA}} & \textbf{Correlation} \\ \hline
\multicolumn{1}{|l|}{\textbf{Silhouette Score}}       & \multicolumn{1}{c|}{0.47574}                    & \multicolumn{1}{c|}{0.51564}                 & 0.48323                    & \multicolumn{1}{c|}{0.73409}                    & \multicolumn{1}{c|}{0.68170}                 & 0.72274                    \\ \hline
\multicolumn{1}{|l|}{\textbf{Davies Bouldin Index}}   & \multicolumn{1}{c|}{0.55996}                    & \multicolumn{1}{c|}{0.49664}                 & 0.54936                    & \multicolumn{1}{c|}{0.38857}                    & \multicolumn{1}{c|}{0.42842}                 & 0.40737                    \\ \hline
\multicolumn{1}{|l|}{\textbf{Calinski Habrasz Index}} & \multicolumn{1}{c|}{72.97150}                   & \multicolumn{1}{c|}{75.68820}                & 75.72814                   & \multicolumn{1}{c|}{81.62642}                   & \multicolumn{1}{c|}{78.73628}                & 79.16408                   \\ \hline
\end{tabular}
}
\end{table*} 
Several studies have used fitness data to model its wearer behaviour for fitness purposes \cite{ni2019modeling}. Hence,  we propose to \textit{use fitness data to recognize users' activities for energy harvesting purposes}.  Even though people may do the same activity, e.g., walking, they may differ in activity intensity level, which may harvest different amounts of energy \cite{khalifa2017harke}. Therefore, we propose an activity-based profiling framework to identify macro profiles that represent the major activity behaviour of IoT users.  These profiles will be used to build an energy harvesting model to estimate the harvested energy based on the users' activity and the used energy harvester.\looseness=-1

\vspace{-5pt}
\section{Profiling Framework}
\begin{figure}[!t]
    \centering
    \setlength{\abovecaptionskip}{5pt}
    \setlength{\belowcaptionskip}{-15pt}
    \includegraphics[width=0.9\linewidth]{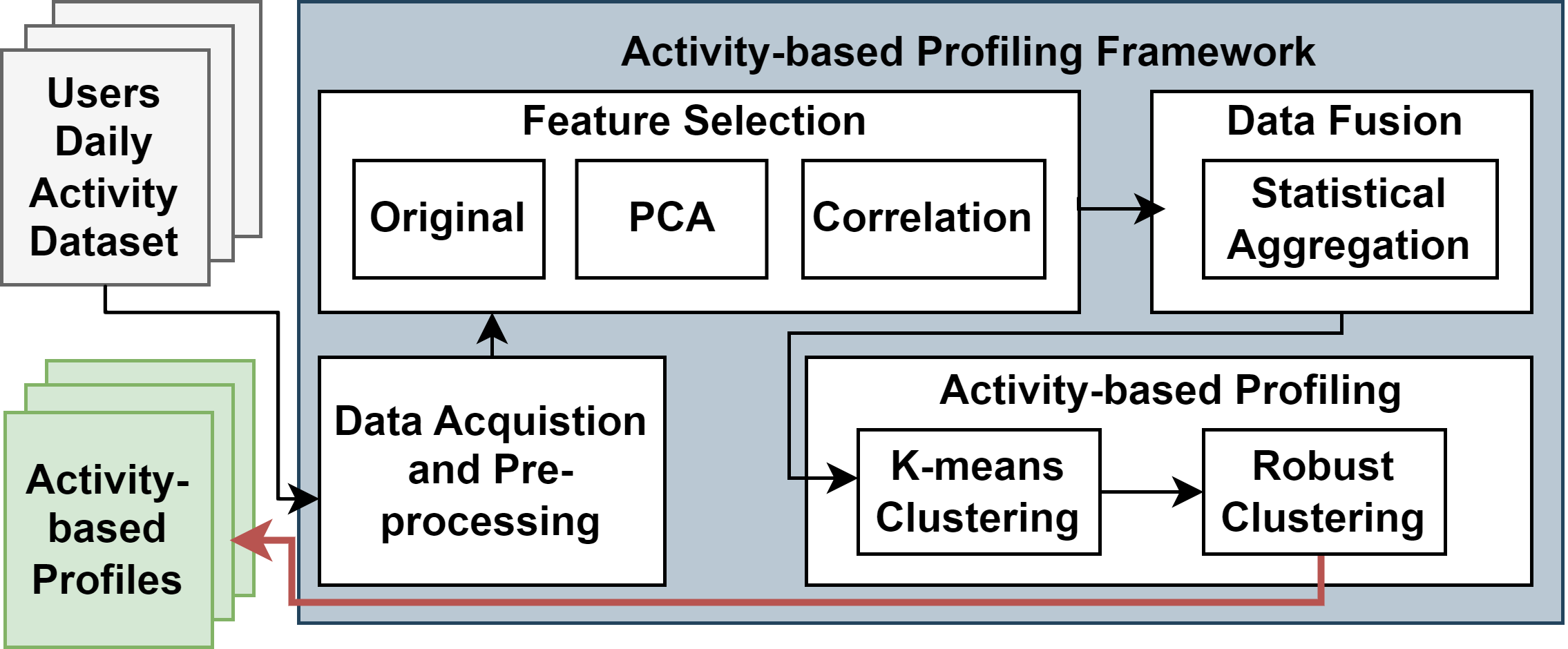}
    \caption{Activity-Based Profiling System Architecture}
    \label{framework}
\end{figure}

The activity-based  profiling framework aims to cluster  users according to their daily activity records. The  framework consists of four phases: Data Acquisition and Pre-processing, Feature Selection, Data Fusion and Activity-based  Profiling (See Fig. \ref{framework}). In what follows we discuss each phase:\looseness=-1

\begin{enumerate}[noitemsep,nosep,leftmargin=0pt,labelsep=0pt,itemindent=8pt]
\item \textbf{Data Acquisition and Pre-processing:}
During this phase, users' health-related information and daily activity data are obtained from their fitness tracker. We used a public Fitbit dataset consisting of 940 records of 33 users' daily activities over a 2-month period in 2016\footnote{kaggle.com/arashnic/fitbit}. As the dataset does not have bio-related data for all users, only activity-related data is used. Also, three features are removed from the dataset: ActivityDate', TrackerDistance', and LoggedActivityDistance' because they are exact duplicates for other attributes.\looseness=-1

\item{\textbf{Feature Selection:}}\label{FS} Feature selection
identifies the most representative features of an IoT user's daily activity. The following three approaches were applied separately to the original cleaned dataset, resulting in three processed datasets:\looseness=-1

\begin{enumerate}[noitemsep,nosep,leftmargin=10pt,labelsep=0pt,itemindent=10pt]
\item \textbf{Original module:} This module contains all the features  of the pre-processed dataset.\looseness=-1
\item \textbf{PCA module:} We  apply the Principal Component Analysis (PCA) technique on the original dataset to transform all the attributes into only three components.\looseness=-1
\item \textbf{Correlation module:}
This module uses a feature-based heat map to select features based on their correlation, removing highly positively or negatively correlated features.\looseness=-1
\end{enumerate}

\item{\textbf{Data Fusion}} Clustering users into groups is not applicable to our fitness dataset as it has multiple activity records for each user. Therefore, data fusion is needed to aggregate the records of each user with minimal information loss. Specifically, we used the statistical values (maximum, minimum, range, standard deviation, mean, and median) to capture user characteristics. These statistics are computed for each user in the three processed datasets from step (\ref{FS}).\looseness=-1 
\item{\textbf{Activity-based Profiling}}\label{clustering} This phase includes applying \textit{K-means} clustering to users' aggregated records, followed by using the resulting clusters in a robust clustering approach.\looseness=-1  

\begin{enumerate}[noitemsep,nosep,leftmargin=10pt,labelsep=0pt,itemindent=10pt]
\item{\textbf{K-means Clustering}} K-means clustering is applied to each of the datasets obtained earlier. We used the \textit{elbow} method to determine the optimal number of clusters for each dataset. As a result, each dataset produced four clusters.\looseness=-1  
\item{\textbf{Robust Clustering}}
As aforementioned, the outcome of  the clustering was four clusters for each dataset, but with varying cluster sizes (number of users).  To address this, we define \textit{robust clustering} by identifying common users across the same clusters in all three sets. Users who are clustered together in all three K-means sets are defined as one robust cluster. Users with different clustering neighbours in the original three K-means sets are defined as separate robust clusters. This produces seven robust clusters/profiles, each representing an activity-based macro profile for a group of IoT users with common behaviour, i.e., activity-based behaviour. A harvested-energy estimation machine-learning model may be used to associate each profile with an estimated energy amount based on its characteristics, using a user's daily records, profile, and type of harvester worn. The energy harvesting estimation model will be investigated in the future.\looseness=-1
\end{enumerate}
\end{enumerate}  
\vspace{-5pt}
\section{Evaluation Metrics}
The performance of our robust clustering algorithm is evaluated using three metrics: Silhouette Score ($SS$), Davies-Bouldin Index ($DBI$), and Calinski-Habrasz Index ($CHI$)\cite{baarsch2012investigation}. We compare the performance of the robust clustering algorithm with the K-means clustering  applied to each feature selection module(See step\ref{clustering}). To calculate the  metrics, each set of features is used with the robust clustering. The results of the metrics are shown in Table \ref{resultstable}. The $SS$ ranges from -1 to 1, with a score closer to 1 indicating denser and more distinct clusters. The robust clustering approach outperforms K-means clustering. The original module achieved the highest $SS$. In contrast, a lower $DBI$ represents a better separation among clusters. The robust clustering with all modules has a $DBI$ than k-means clustering, with the original module having the lowest index. A higher $CHI$ indicates better clustering performance, and the robust clustering approach with all modules outperforms K-means clustering, with the original module achieving the highest index.\looseness=-1
\vspace{-5pt}
\begin{acks}
This research was partly made possible by E180100158 grant from the Australian Research Council. The statements made herein are solely the responsibility of the authors.
\end{acks}
\vspace{-5pt}
\bibliographystyle{unsrt} 
\bibliography{main}

\begin{thebibliography}{10}

\bibitem{lakhdari2020Vision}
Abdallah Lakhdari et~al.
\newblock Crowdsharing wireless energy services.
\newblock In {\em CIC}, pages 18--24. IEEE, 2020.

\bibitem{abusafia2022services}
Amani Abusafia et~al.
\newblock Service-based wireless energy crowdsourcing.
\newblock In {\em ICSOC}. Springer, 2022.

\bibitem{sandhu2020phd}
Muhammad~Moid Sandhu.
\newblock Phd forum abstract: Energy harvesting based sensing for the
  batteryless iot.
\newblock In {\em IPSN}, pages 373--374. IEEE, 2020.

\bibitem{gorlatova2015movers}
Maria Gorlatova et~al.
\newblock Movers and shakers: Kinetic energy harvesting for the internet of
  things.
\newblock {\em Jrnl on Selected Areas in Comm.}, 33(8):1624--1639, 2015.

\bibitem{abusafia2022maximizing}
Amani Abusafia and et~al.
\newblock Maximizing consumer satisfaction of iot energy services.
\newblock In {\em ICSOC}. Springer, 2022.

\bibitem{Amani2022QoE}
Amani Abusafia and et~al.
\newblock Quality of experience optimization in iot energy services.
\newblock In {\em ICWS}. IEEE, 2022.

\bibitem{yang2023Monitoring}
Pengwei Yang et~al.
\newblock Monitoring efficiency of iot wireless charging.
\newblock In {\em IEEE Percom}, 2023.

\bibitem{yao2022wireless}
Jessica Yao et~al.
\newblock Wireless iot energy sharing platform.
\newblock {\em IEEE Percom}, 2022.

\bibitem{yang2022towards}
Pengwei Yang et~al.
\newblock Towards peer-to-peer sharing of wireless energy services.
\newblock In {\em ICSOC}. Springer, 2022.

\bibitem{khalifa2017harke}
Sara Khalifa et~al.
\newblock Harke: Human activity recognition from kinetic energy harvesting data
  in wearable devices.
\newblock {\em TMC}, 2017.

\bibitem{khalifa2015energy}
Sara Khalifa et~al.
\newblock Energy-harvesting wearables for activity-aware services.
\newblock {\em IEEE internet computing}, 19(5):8--16, 2015.

\bibitem{ni2019modeling}
Jianmo Ni et~al.
\newblock Modeling heart rate and activity data for personalized fitness
  recommendation.
\newblock In {\em www}, pages 1343--1353. ACM, 2019.

\bibitem{baarsch2012investigation}
Jonathan Baarsch et~al.
\newblock Investigation of internal validity measures for k-means clustering.
\newblock In {\em IMECS}, volume~1, pages 14--16. sn, 2012.

\end{thebibliography}
\end{document}